\documentclass[a4paper,11pt]{article}
\usepackage{pos}

\title{Modeling Particle Acceleration and MWL Emission of a PeVatron Microquasar V4641 Sgr}
\ShortTitle{Modeling PeVatron Microquasar V4641 Sgr}

\author*[a]{Anton Dmytriiev}
\author[a]{Frans van der Merwe}
\author[a]{Markus B\"ottcher}

\affiliation[a]{Centre for Space Research, \\ North-West University, Potchefstroom, 2520, South Africa}

\emailAdd{amdmame@gmail.com}
\emailAdd{frans.h.vandermerwe@gmail.com}
\emailAdd{markus.bottcher@nwu.ac.za}

\abstract{The Large High Altitude Air Shower Observatory (LHAASO) has recently reported five Galactic microquasars as Ultra-High-Energy (UHE) $\gamma$-ray emitters ($>100$ TeV). Among these sources, the microquasar V4641 Sgr exhibits $\gamma$-ray emission up to $\sim$0.8 PeV, requiring the acceleration of particles to multi-PeV energies, as well as the hardest UHE spectrum. The mechanisms behind particle acceleration to such energies are not well understood. Furthermore, the limited multi-wavelength (MWL) information on this source appears contradictory, further complicating interpretation and suggesting that V4641 Sgr may represent a particularly unusual case. In this work, we present a detailed physical model of V4641 Sgr that combines first-principles simulations of stochastic (turbulent) particle acceleration with MWL emission modeling. We adopt a leptonic scenario in which electrons are accelerated via the second-order Fermi process driven by relativistic strong turbulence ($\delta B/B \sim 1$). The particle energization is simulated using a dedicated Monte Carlo framework \texttt{STRIPE} that incorporates the effects of intermittent energy gains and radiative losses. The resulting accelerated electrons produce UHE $\gamma$-rays through inverse Compton scattering on both the cosmic microwave background (CMB) and the interstellar radiation fields (ISRF). Our model is capable of reproducing key observational characteristics of the system, including particle acceleration to energies of tens of PeV, as well as the TeV-PeV $\gamma$-ray spectrum and the hard spectral index measured by LHAASO. Nonetheless, several aspects remain unresolved, highlighting the need for deeper observational coverage and further theoretical refinement.}

\FullConference{High Energy Astrophysics in Southern Africa (HEASA2025)\\
16-20 September, 2025\\
University of Johannesburg, South Africa\\}

\begin{document}
\maketitle

\section{Introduction}

Microquasars are binary systems consisting of a compact object (a neutron star or a stellar-mass black hole) accreting matter from a companion star, often accompanied by plasma outflows or relativistic jets. These systems offer a unique opportunity to study jet launching and particle acceleration processes within our own Galaxy. Recently, the Large High Altitude Air Shower Observatory (LHAASO) detected five Galactic microquasars in the Ultra-High-Energy (UHE) $\gamma$-ray band ($E_\gamma > 100$ TeV), establishing that, under favorable conditions, microquasars are capable of accelerating particles up to PeV energies. Among these, V4641 Sgr stands out as the most extreme case, exhibiting the hardest UHE $\gamma$-ray spectrum ($\alpha = 2.67 \pm 0.27$) extending up to $\sim$0.8 PeV \cite{lhaaso}, indicating an exceptionally efficient acceleration process operating in this source.

The V4641 Sgr system consists of a black hole with mass $M_{\rm BH} \simeq 6.4 \ M_\odot$ and a B-type companion star in a $\sim$2.8-day orbit, located at a distance of $\sim$6.2 kpc \cite{macdonald2014}. The system also launches a relativistic jet closely aligned with the line of sight, showing superluminal motions and being one of the fastest in the Galaxy \cite{orosz2001}. High-energy observations have revealed extended UHE $\gamma$-ray emission spatially associated with V4641 Sgr. The High Altitude Water Cherenkov Experiment (HAWC) Observatory first reported a pair of elongated $\gamma$-ray ``bubbles'' with a total extension of $\sim$80 pc and a remarkably hard spectrum (photon index $\sim$2.2) detected up to $\sim$200 TeV \cite{hawc}. The LHAASO observations later confirmed this extended emission, while detecting the source up to $\sim$800 TeV \cite{lhaaso}. Meanwhile, H.E.S.S.\ provided a more detailed view of the TeV morphology, revealing a pair of jet-like structures stretching over $\sim$100 pc in total, and measured the spectrum in the range from below 1 TeV up to $\sim$30 TeV \cite{hess}. In contrast, {\it Fermi}-LAT did not detect the source in the GeV range, placing only upper limits on its flux. Interestingly, the direction of the $\gamma$-ray ``bubbles'' is misaligned with respect to the radio jet -- instead, the ``bubbles'' appear oriented nearly perpendicular to the binary's accretion disk, at a viewing angle $i_{\rm b} \simeq 72^\circ$ with respect to the line of sight \cite{hawc}.

At other wavelengths, the XRISM mission detected V4641 Sgr in X-rays in September 2024, revealing spatially extended emission on a scale of $\sim$30 pc, appearing by a factor of a few smaller than that in TeV-PeV band \cite{xrism}. Although the short exposure prevented precise spectral characterization, a useful flux estimate in the keV range was obtained.

The origin of the UHE $\gamma$-ray emission from V4641 Sgr remains uncertain. A hadronic scenario proposed by \cite{neronov2025} interprets the TeV–PeV bubbles as being produced by cosmic rays escaping from the microquasar and interacting with the interstellar medium (ISM). In this model, particles accelerated in the compact radio jet (spanning over the binary scales) escape into the ISM at the jet termination region and stream along ordered Galactic magnetic field lines, explaining the observed misalignment between the radio and $\gamma$-ray structures. This scenario also predicts a substantial neutrino flux. Alternatively, a leptonic model by \cite{wan2025} attributes the UHE $\gamma$-rays to inverse Compton (IC) scattering on the Cosmic Microwave Background (CMB) and interstellar radiation fields (ISRF), with electrons accelerated via shear acceleration within the extended ``bubbles''. Both models successfully reproduce the observed UHE $\gamma$-ray spectra.

In the present work, we investigate a leptonic scenario akin to \cite{wan2025}, but in which electrons are energized through the second-order Fermi acceleration process driven by strong ($\delta B/B \sim 1$) relativistic turbulence. We demonstrate that such a mechanism naturally produces sufficiently hard particle spectra and efficiently accelerates electrons up to tens of PeV, while also reproducing the observed TeV–PeV $\gamma$-ray emission of V4641 Sgr.

\section{The model}

\subsection{General picture and formation of $\gamma$-ray bubbles}


The formation of the large-scale, stable $\gamma$-ray ``bubbles'' extending up to $\sim$100 pc from the microquasar, may be explained by the scenario proposed by \cite{neronov2025}. In this framework, the compact, relativistic Blandford–Znajek (BZ) \cite{bz} jet launched along the black hole spin axis interacts with the ISM at the jet termination region. High-energy particles entrained in the jet escape into the surrounding medium and subsequently stream along large-scale Galactic magnetic field lines. This process naturally produces extended, magnetically guided $\gamma$-ray structures whose orientation may differ from that of the small-scale radio jet, and can also explain the longevity and apparent stability of the observed TeV-PeV structures. We also propose here a second, more speculative possibility, that involves a large-scale Blandford–Payne (BP) \cite{bp} type outflow, magnetocentrifugally launched from the outer regions of the accretion disk and collimated by magnetic hoop stresses and external pressure gradients. These processes can naturally guide the outflow into a direction nearly perpendicular to the accretion disk, explaining the observed orientation of the large-scale $\gamma$-ray structures. However, the BP scenario requires long-term stability of the disk geometry, which may be difficult to satisfy in the misaligned \cite{hawc} V4641 Sgr system.

In both scenarios, the extended $\gamma$-ray bubbles provide an environment where strong magnetohydrodynamic (MHD) turbulence can efficiently re-accelerate the injected electrons. In the picture by \cite{neronov2025}, turbulence is expected to form as the relativistic (BZ) radio jet slams into the surrounding ISM at its termination region. This interaction drives velocity shears, magnetic field distortions and/or shocks, which efficiently convert bulk kinetic and magnetic energy into turbulent fluctuations, which may be further amplified by magnetic reconnection within the tangled field, as well as streaming instabilities, generating turbulence across a very extended distance. Alternatively, in the BP–type disk wind picture, strong turbulence could be generated due to recollimation shocks, as well as kink modes and reconnection, as the outflow collimates and interacts with the external medium. In both scenarios, because the ordered magnetic field in the ISM is relatively weak (a few $\mu$G), even moderate levels of magnetic or kinetic perturbations can produce large relative fluctuations, yielding a regime with $\delta B/B \sim 1$. Moreover, the expansion and diffusion of plasma into $\sim$100 pc-scale volumes implies a very low particle density, leading to a magnetization parameter $\sigma \sim 1$ and consequently relativistic turbulence, with Alfvén velocities $v_{\rm A} = c \ \sqrt{\sigma/(1+\sigma)} \sim c$. The combination of the above-mentioned processes is expected to sustain the bubbles as magnetically confined, turbulent reservoirs where particles are trapped and stochastically re-energized over long time-scales, accounting for the observed stability of the $\gamma$-ray emission.

\subsection{Particle acceleration model}
\label{sec:acceleration}

In the regime of strong ($\delta B/B \sim 1$) and relativistic ($v_{\rm A} \sim c$) turbulence, the quasi-linear theory (QLT) breaks down, and a more complex particle acceleration framework is necessary. Recent MHD and Particle-in-cell (PIC) simulations for these conditions revealed a range of non-resonant phenomena, including particle interactions with non-trivial velocity structures, enhanced acceleration in localized regions of space, etc. Particle energization in this case is dominated not by regular MHD wave–particle interactions, but by sharp bends of magnetic field lines and compression modes perpendicular to the field. These effects make particle energization highly intermittent, especially on small spatial scales \cite{nb2021}. As a result, the acceleration process can no longer be described with a universal diffusion coefficient, and particle transport significantly deviates from the standard Fokker-Planck equation framework. Simulations further demonstrate that particle acceleration can reach very high efficiencies in the $\delta B/B \sim 1$ regimes.  

To model particle acceleration in the regime of strong relativistic turbulence, we adopt the framework by \cite{lemoine2022}, which connects the rate of particle energization to the gradients of the velocity of magnetic field lines. The statistics of these velocity gradients is captured via a multifractal description of turbulence intermittency. The key parameters of this model are the magnetic field $B$, the turbulence coherence length $l_c$, and the effective turbulent Alfvén velocity $\beta_a = \beta_A \ (\delta B/B)$, where $\beta_{\rm A} = \sqrt{\sigma/(1+\sigma)}$ and $\delta B/B = \delta B / \sqrt{B_0^2 + \delta B^2} \leq 1$, with $B_0$ denoting the ordered component of the magnetic field. Within this formalism, the rate of change of the particle momentum in the scattering center frame, is described by a continuous-time random walk (CTRW) where particles interact with magnetic field line velocity gradients \(\Gamma_{l_g}\) coarse grained on scales comparable to the particle's orbit \(l_g \sim 2\pi r_g\), with \(r_g = pc/eB\) being the particle gyroradius, and with momentum evolving as \(\dot{p}=p\Gamma_{l_g}\). The CTRW is then defined by a joint probability for the particle momentum to jump from \(p'\) to \(p\) (with \(\Delta p =p-p'\)) in a time \(\Delta t = t-t'\), with \(\Delta t\) and \(\Delta p\) considered random variables. The interaction time \(\Delta t\) is assumed to be exponentially distributed with the mean value \(\tau=l_g(p')/c\), while the momentum gain (or loss) is distributed as \(\text{PDF}(\Delta \ln p) \sim \text{PDF}(\Gamma_{l_g}) \Delta t\). The probability distribution function capturing the statistics of velocity gradients \(\text{PDF}(\Gamma_{l_g}) \equiv \mathrm{p}_{\Gamma_{l_g}}\) is approximated by a broken powerlaw following \cite{lemoine2022}, with its characteristic parameters strongly depending on the spatial scale $l_g$. In each interaction, a particle can (at random) either gain or lose energy (governed by relative probabilities), with the \(\mathrm{p}_{\Gamma_{l_g}}\) of negative interactions having a smaller characteristic width and thus lower relative probability. The model also accounts for particles escaping the (local; small-scale) velocity gradient accelerating them, as well as for the decoupling of particles from the turbulence at \(l_g \sim l_c\). We further extend the framework to include synchrotron and inverse Compton cooling losses suffered by the electrons (assuming the ultra-relativistic regime with Lorentz factors \(\gamma \gg 1\)). We also incorporate particle escape from the entire (large-scale) acceleration region, with the escape time-scale for an individual particle being exponentially distributed around a characteristic \(t_{\rm esc}\).  Finally, particle momenta (electron spectra) simulated in the scattering center frame are transformed to the lab frame following the procedure in \cite{lemoine2022}.

For the purpose of modeling the acceleration, we developed a dedicated numerical code \texttt{STRIPE} (Strong-Turbulence Relativistic Intermittent Particle Energization), implementing the aforementioned formalism by \cite{lemoine2022} in an MPI parallelized Monte Carlo code written in C.

We assume continuous injection of pre-accelerated particles into the $\gamma$-ray ``bubbles'', where they subsequently undergo turbulent re-acceleration. The particles are pre-accelerated within the (misaligned) radio jet and likely further energized by turbulence-driven magnetic reconnection in the jet termination region. While the composition of the injected plasma is most likely electron–proton, the protons do not significantly contribute to the observed radiation and are therefore disregarded in this model. Since the details of pre-acceleration are beyond the scope of this work, we approximate the injected population as mono-energetic with Lorentz factor $\gamma_0 = 10^5$. We note that such energies are readily achievable through the aforementioned processes, and, importantly, the resulting electron spectra after turbulent re-acceleration are found to be insensitive to the specific choice of $\gamma_0$. Within the extended bubble volume, the injected electrons are subject to turbulent re-acceleration, radiative (synchrotron and IC) cooling, and escape at an energy-independent time-scale $t_{\rm esc}$. The competition between these processes leads to the formation of an asymptotic steady-state electron distribution, which we simulate using the \texttt{STRIPE} code for a range of physical parameters.

\subsection{Emission model}

We adopt a leptonic emission scenario in which particles within the $\gamma$-ray bubbles produce TeV-PeV $\gamma$-rays via IC scattering of ambient photons of the CMB and ISRF, which for electrons above $\sim$100 TeV proceeds in the transition to the Klein–Nishina (KN) regime. The spectral energy density template of the ISRF is taken from \cite{porter2008}, assuming the case of a maximum metallicity gradient and the distance from the Galactic center of $R = 0$ kpc, which is reasonable given the location of V4641 Sgr within the Galactic bar. The X-ray emission in this framework arises from synchrotron emission of the same accelerated electrons. To compute the synchrotron and IC spectra produced by the steady-state electron spectra obtained in \ref{sec:acceleration}, we employ the microquasar emission module of \cite{dmytriiev2024}, further extended for the present work. Given the relatively large distance to the source ($\approx 6.2$ kpc) and its position within the dense photon environment of the Galactic bar, attenuation of UHE $\gamma$-rays through $\gamma$–$\gamma$ pair production on CMB and ISRF photons becomes significant. To account for this effect, we use the attenuation curves calculated by \cite{zhangguo2024} for a set of Galactic microquasars, applying the specific curve for V4641 Sgr to compute the observed absorbed UHE $\gamma$-ray flux.

We assume a cylindrical geometry of the ``bubbles'', with two identical oppositely-directed outflow components moving with a certain bulk Lorentz factor $\Gamma_{\rm b}$. The approaching bubble is viewed at an angle $i_{\rm b} = 72^{\circ}$ \cite{hawc}. The employed emission code readily accounts for the Doppler-boosted radiation from both approaching and receding outflow components. For each bubble, we fix the height (length) at $H = 50$ pc (in the lab frame; relativistically transformed to the comoving frame) and adopt a radius $R_{\rm b}$ in a narrow range of 5 -- 8 pc, based on preliminary results by H.E.S.S. \cite{hess}. The turbulence coherence length is linked to the radius of the bubble and is set to $l_c = 0.3 \times R_{\rm b}$, an empirically motivated ratio commonly used in PIC simulations (M.~Lemoine, private comm.).

\section{Results}

\begin{table}
    \centering
    \begin{tabular}{|c|c|c|c|c|c|c|c|}
    \hline
        $B$ ($\mu$G) & $l_c$ (pc) & $\beta_a$ & $t_{\rm esc}$ ($l_c/c$) & $n_{\rm e}$ (cm$^{-3}$) & $\Gamma_{\rm b}$ & $\delta B/B$ & $\sigma$   \\
        \hline
        3 & 1.5 & 0.41 & 6 & $5 \times 10^{-14}$ & 1.5 & 0.64 & 0.7 \\
        \hline
    \end{tabular}
    \caption{Best fit parameters of the turbulent particle acceleration model combined with leptonic emission scenario, applied to TeV-PeV $\gamma$-ray spectrum of V4641 Sgr, along with the available X-ray flux constraints. The last two parameters ($\delta B/B$, $\sigma$) are derived quantities, rather than fit.}
    \label{tab:params}
\end{table}

\begin{figure}
    \centering
    \includegraphics[width=0.45\linewidth]{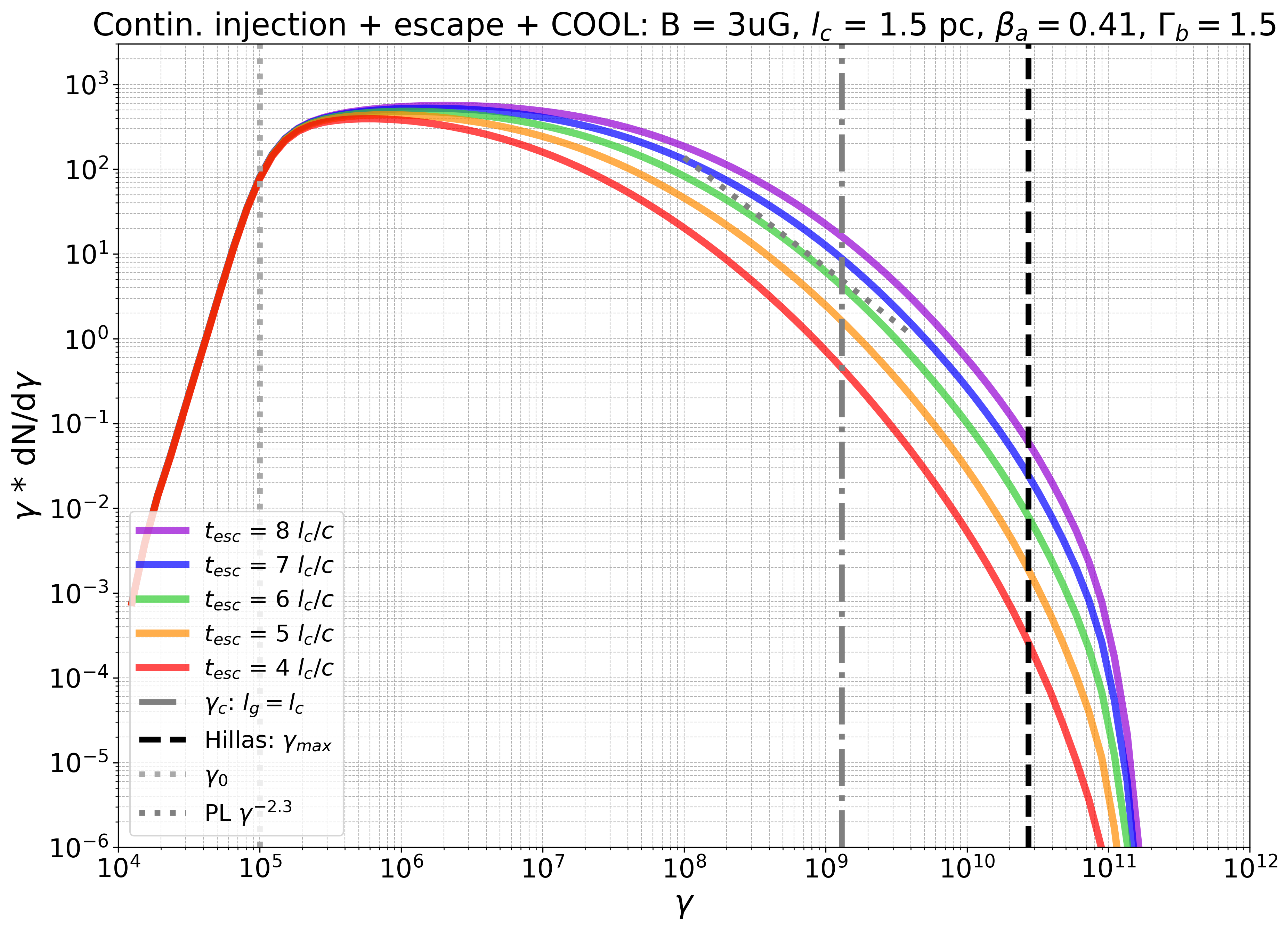} \includegraphics[width=0.66\linewidth]{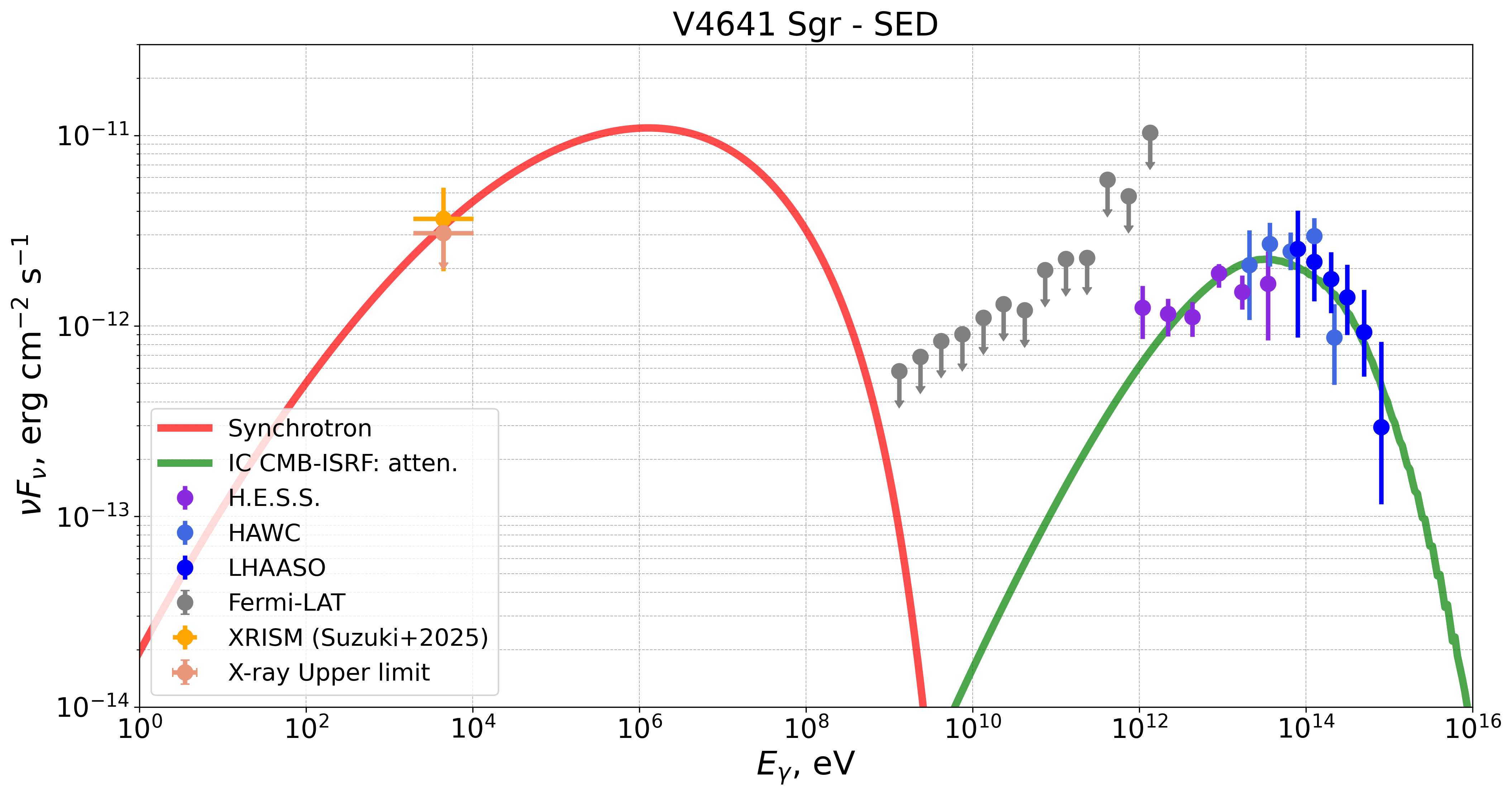}
    \caption{Top: simulated electron spectra resulting from strong turbulent re-acceleration for different $t_{\rm esc}$. The successful SED fit is achieved using the electron spectrum with $t_{\rm esc} = 6 \ l_c/c$. Bottom: composite SED fit of TeV-PeV $\gamma$-ray and X-ray data of V4641 Sgr. The H.E.S.S., HAWC, LHAASO and {\it Fermi}-LAT data is taken from \cite{neronov2025}, X-ray flux measurement from \cite{xrism}, and the X-ray flux upper limit is derived based on \cite{wan2025}.}
    \label{fig:sed_modeling}
\end{figure}

We vary the turbulent Alfvén velocity $\beta_a$, the turbulence coherence length $l_c$ (within the range corresponding to the bubble radius $5 \ \text{pc} \leq R_{\rm b} \leq 8 \ \text{pc}$), the bulk Lorentz factor $\Gamma_{\rm b}$, the electron number density $n_{\rm e}$, and the escape time-scale $t_{\rm esc}$. As a result, the turbulent re-acceleration scenario combined with the leptonic IC emission model, successfully reproduces the observed TeV-PeV $\gamma$-ray spectrum of V4641 Sgr. The best-fit parameters are listed in Table~\ref{tab:params}. We find $l_c = 1.5$ pc, implying $R_{\rm b} = 5$ pc. The underlying electron spectra (for different values of $t_{\rm esc}$) are depicted in the top panel of Fig.~\ref{fig:sed_modeling}, while the broadband SED fit is presented in the bottom panel of Fig.~\ref{fig:sed_modeling}. From the obtained parameters, we derive the steady-state magnetization parameter $\sigma = B^2/(4\pi w) \approx 0.7$, where $w = (4/3) \epsilon$ is the enthalpy density and $\epsilon = \langle \gamma \rangle n_{\rm e} m_{\rm e} c^2$ is the electron energy density, with $\langle \gamma \rangle$ denoting the Lorentz factor averaged over the steady-state electron spectrum. We also infer $\delta B/B = \beta_a/\beta_{\rm A} = \beta_a \ \sqrt{(1+\sigma)/\sigma} \approx 0.64$. 

The steep low-energy decline in the {\it Fermi}–LAT band arises naturally in the turbulent acceleration framework: short acceleration time-scales at low energies, combined with intermittent large upward jumps, produce a concave-down electron spectrum with a hard low-energy slope. While such behavior is difficult to obtain with standard shock acceleration, which typically yields softer spectra, other mechanisms capable of producing hard electron spectra (e.g.\ magnetic reconnection) cannot yet be ruled out. Current Fermi–LAT upper limits still allow somewhat softer low-energy slopes. Deeper LAT observations would help to further constrain the low-energy particle population.

The model parameters exhibit a degree of degeneracy, and the computational cost of \texttt{STRIPE} prevents a full parameter-space exploration; the inferred values should therefore be regarded as order-of-magnitude estimates. Sparse tests nevertheless allow some constraints to be identified. The turbulence coherence length, within the physically motivated range 0.1 -- 0.5 $R_{\rm b}$ ($0.5 \ \text{pc} \lesssim l_c \lesssim 4 \ \text{pc}$), still yields acceptable fits. The bulk Lorentz factor is moderately well constrained by the data at $\Gamma_{\rm b} \approx 1.5$. In contrast, the turbulent Alfvén speed $\beta_a$ and escape time-scale $t_{\rm esc}$ remain less precisely determined and exhibit compensatory behavior. Trial runs indicate that values $\beta_a \leq 0.3$ or $\beta_a \geq 0.5$ no longer reproduce the observed SED, with tighter constraints requiring a dedicated, computationally intensive parameter study.

A notable caveat of this model is that the measured extent of the X-ray emission ($\sim$30 pc) is considerably smaller than that observed in UHE $\gamma$-rays \cite{xrism}. This likely points to an additional, distinct population of electrons located closer to the central source. While a full two-zone treatment is beyond the present scope, we use the available X-ray flux as an upper-limit constraint. To avoid overshooting this flux, the magnetic field in the bubble should not exceed $\sim$3 $\mu$G; hence, this value should be rather regarded as an upper bound.

To verify the self-consistency of this scenario, we estimate the relevant characteristic time-scales. For the inferred parameters, the acceleration time-scale is $t_{\rm acc} \sim l_c/c \ \sim $ 5 yr. The (synchrotron plus IC) cooling time-scale for 100 TeV electrons is $t_{\rm cool} \sim 7.7 \times 10^3$ yr, while the escape time-scale is $t_{\rm esc} = 6 \ l_c/c \ \sim$ 30 yr. Since $t_{\rm acc} < t_{\rm esc} < t_{\rm cool}$, electrons can be efficiently energized, reaching energies of tens of PeV. Next, the diffusion length of the highest-energy electrons over the escape time is $r_{\rm diff} \sim \sqrt{2 D t_{\rm esc}}  \sim \sqrt{2 c \ l_c \ t_{\rm esc}} \approx 5$ pc, in good agreement with the derived radius $R_{\rm b}$, thus ensuring overall consistency of the model. Explaining the larger longitudinal extent of the bubbles ($\sim 50$ pc) is more challenging. While the diffusion length computed over the cooling time $t_{\rm cool}$ reaches $\sim 80$ pc, transport along the ordered Galactic magnetic field is still limited by the shorter escape time. Assuming streaming along the bubble axis, the maximal distance before escape, $r_{||} = c t_{\rm esc} = 6 l_c = 9$ pc, remains below the observed size, implying that a longer effective escape time (or slower parallel diffusion) may be required. Finally, based on the derived parameters, we estimate the total kinetic jet power using Eq.~31 in \cite{wan2025}, obtaining $L_{\rm kin} \sim 4.6 \times 10^{32}$ erg s$^{-1}$, well below the Eddington luminosity of V4641 Sgr of $L_{\rm edd} \simeq 1.3 \times 10^{39}$ erg s$^{-1}$.

\section{Conclusions}

Strong ($\delta B/B \sim 1$) relativistic ($\beta_a \sim 1$) turbulence, excited on large spatial scales beyond the radio jet termination region, provides a natural mechanism for accelerating particles in the V4641 Sgr system up to $\sim$10 -- 50 PeV. IC upscattering of CMB and ISRF photons by the accelerated electrons is able to explain the observed TeV–PeV $\gamma$-ray spectrum, including the hard spectral slope and the extension of the LHAASO spectrum up to $\sim$800 TeV. Constraints from the X-ray data imply an upper limit on the magnetic field of $\sim$3 $\mu$G.

\section{Acknowledgments}

We thank M.~Lemoine and L.~Comisso for useful discussions. The authors acknowledge support from the Department of Science, Technology and Innovation, and the National Research Foundation of South Africa through the South African Gamma-Ray Astronomy Programme (SA-GAMMA).


{\small 
\begin{thebibliography}{99}

\bibitem{lhaaso} LHAASO Collaboration. 2024, e-print arXiv:2410.08988. doi:10.48550/arXiv.2410.08988

\bibitem{macdonald2014} MacDonald, R. K. D., Bailyn, C. D., Buxton, M. 2014, ApJ, 784, 1. doi:10.1088/0004-637X/784/1/2

\bibitem{orosz2001} Orosz, J. A., Kuulkers, E., van der Klis, M. et al. 2001, ApJ, 555, 1. doi:10.1086/321442

\bibitem{hawc} Alfaro, R. et al. 2024, Nature, 634, 8034.
doi:10.1038/s41586-024-07995-9 

\bibitem{hess} H.E.S.S. talks at multiple conferences: Gamma-2024, VGGRS VII

\bibitem{xrism} Suzuki, H., Tsuji, N., Kanemaru, Y. et al. 2025, ApJL, 978, 2. doi:10.3847/2041-8213/ad9d11 

\bibitem{neronov2025} Neronov, A., Oikonomou, F., Semikoz, D. 2025, PhysRev D, 111, 10. doi:10.1103/PhysRevD.111.103025
 
\bibitem{wan2025} Wan, S.-Y., Wang, J.-s., Liu, R.-Y. 2025, e-print arXiv:2507.02763

\bibitem{bz} Blandford, R. D., Znajek, R. L. 1977, MNRAS, 179, 433-456. doi:10.1093/mnras/179.3.433

\bibitem{bp} Blandford, R. D., Payne, D. G. 1982, MNRAS, 199, 883-903. doi:10.1093/mnras/199.4.883

\bibitem{nb2021} Nättilä, J., Beloborodov, A. M. 2021, 921, 1. doi:10.3847/1538-4357/ac1c76

\bibitem{lemoine2022} Lemoine, M. 2022, PhRL, 129, 21, doi:10.1103/PhysRevLett.129.215101 

\bibitem{porter2008} Porter, T. A., Moskalenko, I. V., Strong, A. W., Orlando, E., Bouchet, L. 2008, ApJ,  682, 1, doi:10.1086/589615

\bibitem{dmytriiev2024} Dmytriiev, A., Zdziarski, A. A., Malyshev, D., Bosch-Ramon, V., Chernyakova, M. 2024, ApJ, 972, 1. doi:10.3847/1538-4357/ad6440 

\bibitem{zhangguo2024} Zhang, J., Guo, Y. 2024,  e-print arXiv:2409.00477


\end{thebibliography}
}
\end{document}